\documentclass[prl,aps,twocolumn,showpacs,superscriptaddress]{revtex4}

\usepackage{graphicx}
\usepackage{bm,amsmath,amssymb}

\newcommand{\beq}{\begin{equation}}
\newcommand{\eeq}{\end{equation}}
\newcommand{\barr}{\begin{eqnarray}}
\newcommand{\earr}{\end{eqnarray}}
\newcommand{\ket}[1]{\vert#1 \rangle}
\newcommand{\bra}[1]{\langle#1 \vert}

\begin{document}

\title{Simulation of time evolution with the MERA}

\author{Matteo Rizzi}
\affiliation{NEST-INFM $\&$ Scuola Normale Superiore, Piazza
        dei Cavalieri 7, 56126 Pisa, Italy}
\affiliation{Max Planck Institut f\"ur QuantenOptik, Hans Kopfermann Strasse
  1, D-85748 Garching, Germany}
\author{Simone Montangero}
\affiliation{NEST-INFM $\&$ Scuola Normale Superiore, Piazza
        dei Cavalieri 7, 56126 Pisa, Italy}
\author{Guifre Vidal}
\affiliation{School of Physical Sciences. The University of
  Queensland. Brisbane, QLD, 4072, Australia} 

\begin{abstract}
We describe an algorithm to simulate time evolution using the Multi-scale
Entanglement Renormalization Ansatz (MERA) and test it by studying a critical
Ising chain with periodic boundary conditions and with up to $L\approx 10^6$
quantum spins. The cost of a simulation, which scales as $L\log(L)$, is
reduced to $\log(L)$ when the system is invariant under translations. By
simulating an evolution in imaginary time, we compute the ground state of the
system. The errors in the ground state energy display no evident dependence on
the system size. The algorithm can be extended to lattice systems in higher
spatial dimensions.  

\end{abstract}

\date{\today}
 
\maketitle
 
The role of numerical simulations in many branches of physics is becoming more
and more fundamental as the complexity of the systems of interest increases
\cite{dmrgrev}. Recently, the injection of quantum information concepts has
opened up the possibility of significant improvements in our ability to
simulate strongly correlated quantum many-body systems. The entanglement
present in the many-body wave function has been identified as a major limiting
factor in numerical simulations.
Accordingly, a big effort has been made within the quantum information
community to devise new simulation strategies that, building upon the matrix
product state (MPS) \cite{MPS} and the density matrix renormalization group
\cite{DMRG}, include a careful description of entanglement (see
e.g. Refs. \cite{TEBD, vMPS, PEPS, ER, MERA, eisert, briegel}).  
In particular, the notion of \emph{entanglement renormalization} ---the
systematic removal of short-ranged entanglement in the system--- has been put
forward as a means to obtain an efficient \emph{real-space renormalization
  group} (RG) transformation for quantum systems on a lattice
\cite{ER}. Relatedly, the \emph{Multi-scale Entanglement Renormalization
  Ansatz} (MERA) has been proposed as a variational many-body wave function to
describe ground states \cite{MERA}. It has been already demonstrated that the
MERA offers a particularly accurate and compact representation of critical and
non-critical ground states in 1D lattices \cite{ER}. 
However, the existence of an efficient algorithm to systematically compute the
MERA for ground states is not yet demonstrated. 

In this paper we present an algorithm, referred to as the t-MERA algorithm,
to simulate time-evolution with the MERA. 
For simplicity, we describe
and test the approach in a one-dimensional system, namely a critical Ising
chain with periodic boundary conditions. However, the algorithm can be readily
generalized to lattice systems in higher dimensions. The cost of simulating
$L$ spins in an inhomogeneous system scales as $L\log(L)$. We exploit
translational invariance to further reduce this cost to $\log(L)$, allowing us
to accurately address systems of up to $2^{20}\approx 10^{6}$ spins with very
modest computational resources.

The t-MERA algorithm is inspired in the time-evolving block decimation (TEBD)
algorithm for MPS \cite{TEBD}. As in the latter, the tensors in the ansatz are
updated so as to account for the action of a two-body gate acting between two
neighboring lattice sites. However, while in an MPS the update involves only
the tensor immediately close to those sites and is performed with a simple
singular value decomposition \cite{TEBD}, in the case of a MERA the update is
given by a more sophisticated optimization, defined by a fidelity
maximization, as described below.

\begin{figure*}
    \centerline{\includegraphics[scale=0.8]{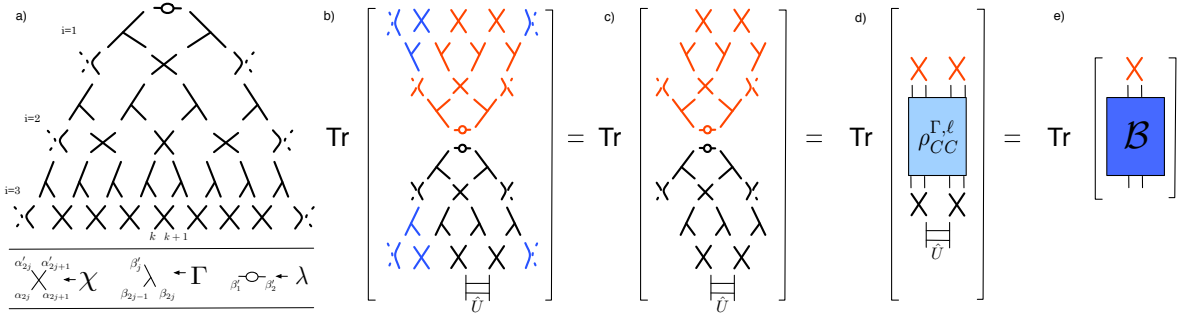}}
    \caption{(Color online) a) MERA tensor structure with periodic boundary
      conditions and $\ell=3$. b) Schematic representation of the fidelity
      (\ref{fid}). Black (red) tensors represents the tensors inside the
      causal cone of $\ket{\psi}$ ($\ket{\tilde \psi}$). Blue  tensors (at sides) are
      outside the causal cone, thus they are contracted for free (c)).
      d-e) Schematic representation of the expressions (\ref{fid1}) and
      (\ref{problem}).} 
    \label{alg}
\end{figure*}

We consider a many body quantum system composed of $L=2^{\ell+1}$ sites, each
of them described by a local Hamiltonian and some nearest neighbor
interactions  $\mathcal{H}_{i,i+1} : \mathbb{C}^{d^2} \rightarrow
\mathbb{C}^{d^2}$ \cite{nnn}. The global Hamiltonian is then 
\beq
\mathcal{H}= \sum_{\langle k\rangle} \mathcal{H}_{k,k+1},
\eeq
where $k=1,L$ with $L+1 \equiv 1$ for periodic boundary conditions. 
Consider the ensemble of wave functions that can be described exactly via a
given MERA structure 
\begin{eqnarray}
&&\mathcal{M} := 
\Big\{ \ket{\psi} \in \mathcal{H}_d^{\otimes L}  \Big|
\ket{\psi} =  \\
&& =\prod_{i=1}^{\ell} \prod_{j=1}^{2^{i}} \lambda_{\beta_1',1}^{\beta_2',1} \hat \chi[j,i] \hat
\Gamma[j,i]   
\delta_{\alpha'_{2j,i}}^{\beta_{2j,i}} 
\delta_{\alpha'_{2j+1,i}}^{\beta_{2j+1,i}} 
\delta_{\alpha_{j,i-1}}^{\beta'_{j,i}}, \Big\} \nonumber
\end{eqnarray}
with
$\hat \chi[j,i]=\chi[i]_{\alpha_{2j},\alpha_{2j+1}}^{\alpha_{2j}',\alpha_{2j+1}'}$
unitary operator $(\hat \chi  \hat \chi^\dagger=\hat\chi^\dagger\hat\chi=1)$,
$\hat \Gamma[j,i]=\Gamma[i]_{\beta_{2j-1},\beta_{2j}}^{\beta_{j}'}$ 
isometry $(\hat\Gamma^\dagger\hat\Gamma=1)$, 
$\sum |\lambda |^2= 1$
and 
$\alpha,\beta,\alpha',=1,\dots, \textrm{min}(m,d^{\ell-i+1})$,
where $i$ counts the tensor network level and $j$ the position on a given
level~\cite{ER,note1}. The MERA structure is  represented in
Figure~\ref{alg}~a.  
As shown in \cite{ER} the tensors $\hat \chi$ can be interpreted as
disentanglers. The parameter $m$ is the dimension of the projected space and
is related with the number of state kept $m_{\mathrm eff}$ in the reduced
density matrix of half system in the Density Matrix Renormalization Group
algorithm \cite{DMRG}. 
Indeed, a simple calculation can show that the two quantities are related
through $m_{\mathrm eff}= m^{2 (log_2 L-1)}$. Notice that the logarithmic
scaling allows to describe the quantum correlations present in 1D critical
chains \cite{logscaling,entrev}.

The simulation of time evolution is achieved by updating the MERA when 
the operator $U=e^{-iHt}$ is applied to the state. When the time is
real, $t \in R$, U is an unitary evolution, whereas when $t$ is imaginary, 
$t = -i\eta$ with $\eta \in R$, $U=e^{-H \eta}$ is an euclidean time 
operator that in the large $t$ limit projects onto the ground state of $H$.
We use the Suzuki-Trotter decomposition at a given order 
to obtain a sequence of two sites operators $U_{k,k+1}$ to be applied to the
initial ansatz $|\psi_0\rangle$ \cite{trotter}. 
The problem is then reduced to the application of the operators $U_{k,k+1}$ to
the MERA and to absorb it, recovering the original structure with the minimum
error. In other words, given a $\ket{\psi} \in \mathcal{M}$ and $U_{k,k+1}$ we
want to find $\ket{\tilde\psi} \in \mathcal{M}$ such that 
\beq
\mathcal{\bar F}= \textrm{max}_{\begin{subarray}{c}\ket{ \tilde\psi} \in
    \mathcal{M} \end{subarray}}^{~~~} |\mathcal{F}| =
 \textrm{max}_{\begin{subarray}{c}\ket{ \tilde\psi} \in
    \mathcal{M} \end{subarray}}^{~~~} \big|\bra{\tilde\psi}  U_{k,k+1}
\ket{\psi}\big|.
\label{fid}
\eeq
To perform such maximization efficiently, we implement a recursive procedure
andan optimization of every element belonging to the causal cone
\cite{causalcone}.  
The maximization is carried out for each tensor separately as no exact method
is known (notice that in the TEBD algorithm this is performed exactly via a
single SVD \cite{TEBD}).  
We first set $\ket{\tilde\psi}=\ket{\psi}$ and we compute the trace fidelity
$\mathcal{F} = Tr(U_{k,k+1} \rho^{\chi,\ell}) $ with $\rho^{\chi,\ell}=
\ket{\psi} \bra{\psi}$ as represented in Figure~\ref{alg}~b. As shown in
\cite{TTN} the trace fidelity is equal to the contraction of the part of the
tensor network inside the causal cone, i.e. the part of the network that can
be influenced by a local operation which, at each level, is composed of
maximum two unitaries $\chi$ or three isometries $\Gamma$ (see Fig.~\ref{alg}
c). Indeed, as shown in the figure, due to the properties of the tensor
involved, the part of the network outside the causal cone is contracted for
free. Thus, to compute $\mathcal{F}$ only the reduced density matrix
$\rho_{CC}^{\chi,\ell}$  of the two involved sites  is needed and the
application of any local operator $U_{k,k+1}$ will result in a modification of
the tensors inside its causal cone.
Notice that the fidelity can be expressed as a function of the reduced density
matrix at the upper level writing explicitly its dependence with the tensors
at the last level, that is 
\beq
\mathcal{F}=Tr(U_{k,k+1} \chi[k,\ell] \chi[k+1,\ell]
\rho_{CC}^{\Gamma,\ell}  \tilde\chi[k,\ell]^\dagger \tilde\chi[k+1,\ell]^\dagger),
\label{fid1}
\eeq
as shown in Figure~\ref{alg}~d.
We first update a single tensor, e.g. $\tilde\chi[k,\ell]^\dagger$, 
contracting all the other tensors obtaining
\beq
\mathcal{F}= Tr(\tilde\chi[k,\ell]^\dagger \mathcal{B}),
\label{problem}
\eeq
as shown in Fig.~\ref{alg} e \cite{noteB}.
The maximum of the fidelity is given by $\tilde\chi[k,\ell]= V$ where $V$ is
the unitary part of the polar decomposition of the matrix $\mathcal{B}= V P$
\cite{thpolar}. We then write the analogous relation (\ref{problem}) related
to the second tensor to be maximized and update $\tilde\chi[k+1,\ell]$. The
maximization can be repeated until convergence is reached. We can now express
explicitly the relation (\ref{fid1}) with its dependence from the isometries
of the last level $\Gamma$s and the updated $\tilde \chi$s 
\beq
\begin{aligned}
\mathcal{F}=Tr(\tilde U_{k,k+1} \Gamma[k,\ell] \Gamma[k+1,\ell]
\Gamma{[k+2,\ell]} \\ \rho_{CC}^{\chi,\ell-1}
\tilde\Gamma[k,\ell]^\dagger \tilde\Gamma[k+1,\ell]^\dagger
\tilde\Gamma[k+2,\ell]^\dagger 
).
\end{aligned}
\label{fid2}
\eeq
with  $\tilde U_{k,k+1} =\tilde\chi[k,\ell]^\dagger 
\tilde\chi[k+1,\ell]^\dagger U_{k,k+1} \chi[k,\ell] \chi[k+1,\ell]$.
To perform the fidelity maximization we repeat the previous operations
optimizing the $\Gamma$s separately, defining at every optimization a new
operator $\mathcal{B}$ and performing its polar decomposition. Notice that in
this case $\mathcal{B}$ is a rectangular matrix and $V$ is an
isometry. Finally the same procedure is repeated for every level of the tensor
structure, until the top is reached. Particular attention is needed to perform
the update of the uppermost $\Gamma$s and the vector $\lambda$. In this case
one can again write the problem to be optimized in the same way as
before. However both $\Gamma$s can be updated together: once computed the
operator $\mathcal{B}$ it can be Schmidt decomposed in two different tensors
each one defining the new $\Gamma[j,1]$s. The singular values obtained define
the new vector $\lambda$. In the case of
euclidean evolution, a renormalization (enforcing $\sum_i
|\lambda_i|^2 =1$) will take in account of the loss of norm of $\ket{\psi'}$
due to the non unitarity of the euclidean operator $\hat U$. 
\begin{figure}
    \centerline{\includegraphics[scale=0.325]{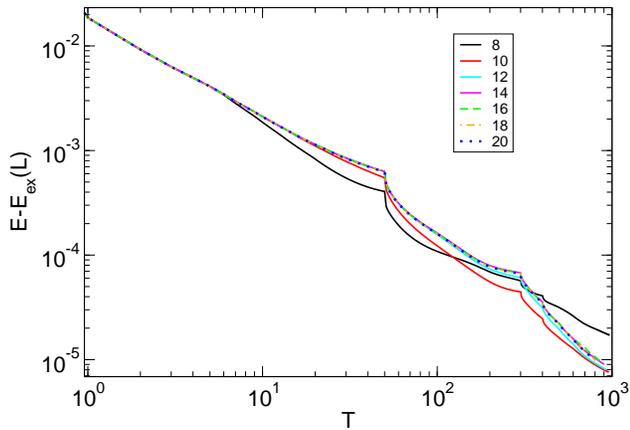}} 
\caption{(Color online) Convergence of the computed ground state energy  $E(L)$ for
  $L=2^\ell,\ell=8,10 \dots 20$ to the exact ground state energy $E_{ex}(L)$
  with    periodic boundary conditions \cite{lieb}. }
\label{conv}
\end{figure}
We have build an approximation of the wave function $U_{k,k+1} \ket{\psi}$ which
maximizes the fidelity $\mathcal{F}$ and $\ket{\tilde\psi} \in
\mathcal{M}$. We then repeat the operations previously described for every
operator $U_{k,k+1}$ and for every Trotter step $\Delta t$ until the desired
convergence is reached.  

In conclusion, we sketch the algorithm scheme for the sake of clarity:
\begin{enumerate}
\item{Decompose the evolution operator in operators that act on nearest
    neighbor physical sites $U_{k,k+1}$ with the Trotter decomposition at
    desired order.} 

\item{For every $U_{k,k+1}$ and for every Trotter step, set $\ket{\psi}
    \rightarrow \ket{\tilde \psi}$ and perform the following actions:}

\begin{enumerate}
For every level $i>1$

\item{Find the optimal $\chi$s obtained by  maximization of the fidelity
    (\ref{fid1}). Replace the old $\chi$s with the new ones, $\tilde \chi$s,
    obtained as unitary part of the polar decomposition of the operator
    $\mathcal{B}$. If necessary repeat the maximization until convergence is
    reached.} 

\item{Repeat step (a) to update the $\Gamma$s maximizing the fidelity
    (\ref{fid2}). Again repeat the process if needed.}\\  
when the uppermost level is reached ($i=1$)
\item{Find the new isometries and the new norm vector via a Schmidt
    decomposition of unitary part of the operator $\mathcal{B}$. 
    If euclidean evolution is simulated, 
    renormalize the vector of the singular values to obtain 
    $\tilde \lambda$. Set $\ket{\tilde \psi} \rightarrow \ket{\psi}$.} 

\end{enumerate}

\item{When desired, perform one and two sites observables measurement
    computing $Tr(\hat O_{k,k'} \rho_{CC}^{\chi,\ell})$ as described in
    \cite{TTN}.} 
\end{enumerate}
The t-MERA algorithm, as described before, requires an impressive limited
amount of resources (e.g. few hours on a laptop for $L=2^{14}$ for the
translational invariant case), polynomial both in memory and time as a
function of the system size and of the desired $m$: given $L$ and $m$ the
memory resources scale as $\mathcal{O}(m^4 L \log L)$ that is the memory
needed to store the tensor structures \cite{ER}. The computational times are
dominated by the tensor contractions needed to compute the operator
$\mathcal{B}$ during the trace fidelity maximization, made of at most $m^9$
operations \cite{TTN}. As for every Trotter step one needs to compute
$\mathcal{O} (L \log L)$ different $\mathcal{B}$ operators (one for each link
and level), the algorithms scales as $\mathcal{O} (m^9 L \log L)$. Even though
the scaling with the projected size $m$ is polynomial, it might still need a
huge amount of computational time for big $m$ due to the high polynomial
scaling power which might result in a limitation of this algorithm
usefulness. However, as we show later on, already with $m=4$ we obtained very
high precisions. More important and differently from previous proposed
algorithms, the projected space size $m$ needed to keep the error constant
does not depend on the system size $L$, allowing to increase the size of the
system up to thousands sites with only a linear-log cost in term of
computational resources. 
Finally, if the Hamiltonian is traslational invariant, one can take advantage
of this symmetry reducing the simulation cost to $O(\log L)$ allowing
impressive system size to be studied as shown in the following. 
\begin{figure}
\centerline{
    \includegraphics[scale=0.325]{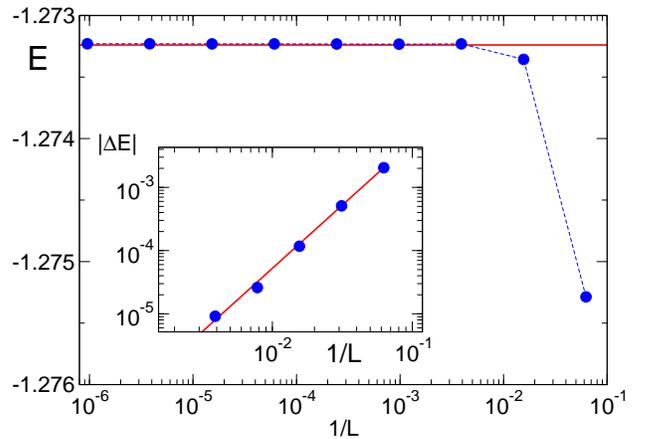}}
    \caption{(color online) Energy $E(L)$ as a function of the system size
      ($L=2^\ell, \ell=8,\dots,20$) for the critical Ising model (blue
      circles), $m=4$, $dt=0.1$ and $T_f=800$. Red full line represents the
      exact thermodinamical limit $E_\infty^{ex}= -4/\pi$. Inset: Difference
      of the computed energy with respect to thermodinamical limit $\Delta E =
      E(L) - E_\infty^{ex}$. The red line represents the exact scaling.} 
    \label{ising}
\end{figure}

\begin{figure}
\centerline{
    \includegraphics[scale=0.325]{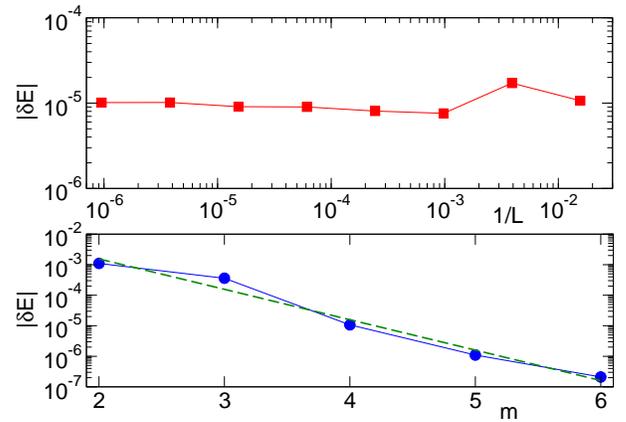}}
    \caption{(color online) A. Error as a function of the system size $L$ of
      the computed energy with respect to the exact one $\delta E = E(L) -
      E_{ex}(L)$, $m=4$, $dt=0.1$ and $T_f=800$ \cite{lieb}.
      B. Error $\delta E$ as a function of $m$ for $L=32$, $dt=0.1$ and
      $T_f=800$. The dashed green line is an exponentail fit.}  
    \label{scale}
\end{figure}

{\it Results:}
We now apply the t-MERA algorithm to the study of the Ising chain ground state
as a benchmark of the precision of the results that can be obtained.
The Ising model is defined as 
\beq
\mathcal{H}= - \sum_{<k>} h \, \sigma_k^z + \sigma_k^x \sigma_{k+1}^x,
\eeq
where $<k>$ describes periodic boundary conditions. 
The model is known to be critical for $h=1$ and it can be solved exactly via
the usual mapping to the fermionic operators \cite{lieb}. In Fig.~\ref{conv}
we plot some typical convergence of the ground state energy as a function of
the imaginary time $T$ for the critical ($h=1$) Ising model with periodic
boundary conditions starting from the completely polarized state $\ket{\psi_0}
= \ket{\uparrow \dots \uparrow}$.  We start with disentanglers set to the
identity and after some time $T_i$ we switch on the optimization procedure for
the $\chi$s as can be seen in Fig.~\ref{conv} where a non smooth behavior is
present. In Figure~\ref{ising} we plot the finite size scaling energy
resulting from our simulations at given final time $T_f$: the resulting energy
is $E(\infty)=1.273229\dots$ with an error with respect to the exact solution
of $\Delta E = 10^{-5}$. As clearly seen in the inset of Fig.~\ref{ising} with
this algorithm we can accurately reproduce the finite size scaling of
properties such as the energy. Finally in Fig.\ref{scale}~A we show the error
$\delta E = E(L) - E_{ex}(L)$ at given time $T_f$ of the computed energy $
E(L)$ with respect to the exact energy for finite size  $E_{ex}(L)$ as a
function of the system size $L$: the error appears to be size
independent. This is the most noticeable feature of the t-MERA algorithm and
it reflects the underlaying MERA tensor structure~\cite{ER}. In
Fig.\ref{scale}~B we show the exponential dependence of the error with respect
to the cut dimension $m$ for $L=32$: Increasing $m$ the error decrease
exponentially while the resources needed by the algorithm (memory and
CPU-time) scale polynomially.
We mention that similar scaling of the errors have been recorded for local and
nearest neighbor observables (errors of order $10^{-3}$ at $T_f=800$) and for
different critical models (data not shown).

In conclusion, the idea introduced here works efficiently on every tensor
network which has a finite size causal cone, that is, we can apply this
algorithm to the proposed extension of 2D MERA \cite{MERA} structures with no
fundamental changes. Moreover, the extensions of the t-MERA algorithm to
include long range interactions and to study open system are also possible
following \cite{MERA, vidalopen}. 
The exponential suppression of the error increasing the cut dimension $m$ in
comparison to the polynomial scaling of the resources also in critical systems
are features that if confirmed in higher dimensionality, candidate this
algorithm for the study of critical systems unaffordable with different
methods. The limitation of this algorithm arises from the computational times
needed, however, the code parallelization can be easily implemeted \cite{wip}.

After the completion of this work MERA tensor structure have been used to
describe ground state in 2D lattice \cite{2DMERA} and topological order
\cite{Topological}. SM and MR thank V.Giovannetti and R. Fazio for very useful
comments and encouragement, I.Latorre for interesting discussions,
M. Montangero and  C. Montangero for useful suggestions regarding the software
development and ackowledge support by EC-FET-EUROSQIP and by Centro di Ricerca
Matematica ``Ennio De Giorgi'' of Scuola Normale Superiore. G.V. acknowledges
finantial support from the Australian Research Council, FF0668731.


\begin{thebibliography}{99}
\bibitem{note1}{For the sake of simplicity we omitted the notation for
  the boundaries of the tensor network which should be taken
  carefully in account}.
\bibitem{causalcone} The causal cone is defined as the part of the tensor
  structure which is reached starting from the physical indexes connected by
  the evolution operator $U_{k,k+1}$ percolating up to the top of the
  following the contracted indexes upwards.
\bibitem{noteB} To simplify the notation we omitted the indexes related to the
  operator $\mathcal{B}$.
\end{thebibliography}
\end{document}